\documentclass{article}
\usepackage{amsmath,amsfonts,latexsym}
\usepackage{amssymb}
\newtheorem{Definition}{Definition}
\newtheorem{lemma}{Lemma}

\newtheorem{proposition}{Proposition}
\newtheorem{corollary}{Corollary}
\usepackage{graphicx}
\graphicspath{%
    {converted_graphics/}% inserted by PCTeX
    {pngs-for-OptRec/}% inserted by PCTeX
}
\begin{document}

\title{Finding Optimal Sinks for Random Walkers in a Network}         % Enter your title between curly braces
\author{Fern Y. Hunt\\
Applied and Computational Mathematics Division\\
National Institute of Standards and Technology\\
Gaithersburg,Maryland 20899}        % Enter your name between curly braces
\date{\today}          % Enter your date or \today between curly braces
\maketitle
%
%%%%%%%%%%%%%%%%%%%%%%%%%%%%%%%%%%%%%%%%%%%%%%%%%%%%%%%%%%%%%%%%%%%%%%%%%%%%%%%%%%%%%%%%%%%%%%%%%%%%%%%%%%%%%%%%%%%%%%%%%%%%%%%%
%
\begin{abstract}
In a model of network communication based on a random walk in an undirected graph, what subset of nodes (subject to constraints on the set size), enables the fastest spread of information? In this paper, we assume the dynamics of spread is described by  a network consensus  process, but to find the most effective seeds we consider the target set of a random walk--the process dual to network consensus spread. Thus an optimal set $A$ minimizes the sum of the expected first hitting times $F(A)$, of random walks that start at nodes outside the set. Identifying such a set is a problem in combinatorial optimization that is probably NP hard. However $F$ has been shown to be a supermodular and non-increasing set function and fortunately some results on optimization of such functions exist.

We introduce a submodular, non-decreasing rank function $\rho$, that permits some comparison between the solution obtained by the classical greedy algorithm and one obtained by our methods. The supermodularity and non-increasing properties of $F$ are used to show that the rank of our solution is at least $(1-\frac{1}{e})$  times the rank of the optimal set. When our approximation has a higher rank than the greedy solution, this can be improved to 
$(1-\frac{1}{e})(1+\chi)$ where $\chi >0$ is a constant. A non-zero lower bound for $\chi$ can be obtained when the curvature and increments of $\rho$ are known. \\

%\keywords{consensus models, random walk, first hitting time, sub(super)modular functions}

\end{abstract}
%%

%%%%%%%%%%%%%%%%%%%%%%%%%%%%%%%%%%%%%%%%%%%%%%%%%%%%%%%%%%%%%%%%%%%%%%%%%%%%%%%%%%%%%%%%%%%%%%%%%%%%%%%%%%%%%%%%%%%%%%%%%%%%%%%%%%%%%
%%
%%
\section{Introduction}\label{S:rwalk}
The study of information spread (or dually consensus) in complex networks has been the subject of intense activity in the past decade (\cite{Olfati}, \cite {Rao}, \cite{Kempe}, \cite{Richardson} \cite{Borgs}) as reserachers study the role  of  distinguished subsets of nodes, such as ``leaders''  in consensus models and ``influential spreaders'' in models of information spread. In particular the research reported in references \cite{Richardson}, \cite{Kempe}, \cite{Borgs} developed methods for obtaining optimal spreaders as determined by some measure of subset performance. 
 
Recall that at each time step of a network consensus process, the function value assigned to a node that is not a seed or leader is updated by averaging the current value with those of its neighbors according to a weighted Laplacian matrix which we take to be the transition matrix of a Markov chain. Seed nodes affect the information values  of its neighbors during updating ,but the value assigned to the seed is unchanged. Over time the value for each node converges to a value dictated by the seed nodes. It is desired to select a set subject to cardinality constraints, that will maximize the rate of convergence. As discussed in \cite{Borkar}, the rate of convergence is determined by the Perron-Frobenius eigenvalue of the averaging matrix restricted to the rows and columns indexed by the set complement of the set of seed nodes. Thus an optimal seed set would minimize this eigenvalue. Rather than solve this problem directly we will seek instead the optimal sink (or target set) of the dual of the network consensus process: the associated random walk.  For a set of nodes, we define  $F(A)$ to be the sum of the expected first hitting times of random walks that start at nodes outside the set. As briefly mentioned in \cite{Borkar} the expected first hitting is related to the eigenvalue so that $F(A)$ is small if and only if the eigenvalue is small. From this point of view $A$ is an optimal if $F(A)$ is minimized over all sets with the same or lesser cardinality.
In the discussion that follows the set of all nodes in the network is denoted by $V$ 
\subsection{Problem Background}\label{S:background}
Borkar, Nair and Sanketh, \cite{Borkar} showed that for subsets
$A \subseteq B \subseteq V \,$  and $ j \in V$, $F(A)-F(A\cup \{j\})\geq F(B)-F(B\cup\{j\})$, that is, $F$ is a supermodular function. Thus $-F$ is submodular so if it is bounded our problem is an instance of submodular maximization, a classic problem in combinatorial optimization. The problem for objective functions other than the one we discuss here has found wide application in many areas of computer science particularly machine learning (see e.g. the tutorials containing introductory material on submodular functions, \cite{Bach}, \cite{Bilmes}). In 1987, Nemhauser, Wolsey and Fisher \cite{NemWolFish} proved the fundamental result that a set constructed by the greedy algorithm for  maximizing a bounded submodular set function has an approximation ratio of $(1-1/e)$. Since then Borgs, Brautbar, Chayes and Lucier \cite{Borgs} and  Sviridenko,Vondrak and Ward  \cite{SvirVonWard}, showed that approximations of comparable or better quality could be obtained very efficiently using different methods. Recently, building on the foundational work of Fujishige \cite{Fujishige}, Bilmes, Iyer and Jegelka presented a framework that unifies many disparate approaches to the optimization theory by exploiting the convex-concave like properties of submodular functions \cite{Iyer1}. Aspects of our method fit into this scheme as we discuss in Section \ref{S:closure}. In \cite{Mikesell}, Mikesell, Kenter and Hicks following on \cite{Hunt1} considered a number of heuristic approaches for solving the problem we discuss here for a number of graphs and compared it with our method. The results (see Figure 6, \cite{Mikesell}) showed that it was no worse than the greedy method and was sometimes better. Understanding the reason for this was another motivation for this work. 

There is a body of closely related work in the area of opinion dynamics where so-called "stubborn" agents can have a profound effect on the outcome of consensus algorithms, random gossip algorithms and similar decentralized protocols.  See for example, \cite{Ameur}, \cite{Pirani} and for work on the optimal placement of stubborn agents in the voter model of communication see  \cite{Acemoglu}.
Research on decentralized protocols appears to have
 originated with Borkar and Vairaya \cite{Borkar2} and Tsitsiklis \cite{Tsit} who discussed
communication in networks with nodes modeled by markov decision processes. 
 
Results of our research are also relevant to the design of algorithms for routing in wireless communication systems when location information is not available \cite{Rao}, \cite{Jadbabaie},   identification of influential individuals in a social network \cite{Kempe} and in sensor placements for efficiently detecting intrusions in computer networks \cite{Krause}. 
In their investigation of a leader-follower network, Clark, Bushnell and Poovendran \cite{Clark}, \cite{Clark2} demonstrated the connection between the rate of convergence to network consensus and a supermodular function closely related to ours. Furthermore they showed that the greedy approximation of the set of optimal leaders produces an approximation that is within $(1-1/e)$ of optimal.

   Given  a connected graph  $G=(V,E)$ with $N$ vertices $V$ and edges $E$, information spreads through the network by a process that is dual to the direction of the random walk (see \cite{Barahona}). An optimal sink is defined in terms of a set function $F$ where for a subset $A \subset V$, $F(A)$ is the sum of mean first arrival times to $A$ by random walkers that start at nodes outside of $A$. If $A$ is an effective target set for the random walks then $F(A)$ is small. Thus the optimal set (subject to a cardinality constraint $K$)  minimizes $F(A)$ subject to $|A| \leq K$,
\begin{equation}\label{E:opt}
\min_{ A \subset V,\\\ |A|\leq K} F(A).
\end{equation}
%%%%%%%%%%%%%%%%%%%%
%----------------------------------------------------------------------------------------------------------------------------------%
Recall that a random walker situated at a node $i \in V$, moves to a neighboring node $j \in V$ in a single discrete time
step with probability, 

\begin{equation}\label{E:probtrans}
Prob\{i \rightarrow j\}= 
\{p(i,j)>0,\quad \mbox{if} \quad (i,j) \in E ,
\quad p(i,j)=0,\quad \mbox{if} \quad(i,j) \notin E\}
%p_{i j}>0 , &\text{if }  (i,j) \in E \\
%p_{i j}=0    &\text{otherwise.}
%
\end{equation}
%%%%%%%%%%%%%%%
%\begin{equation} \label{E:probtrans}
%\setlength{\nulldelimiterspace}{0pt}
%Prob(i,j)=\left\{\begin{IEEEeqnarraybox}[\relax][c]{l's}
%p(i,j), &for $(i,j) \in E$\\
%0  &otherwise
%\end{IEEEeqnarraybox}\right.
%\end{equation}
%%%%%%%%%%%%%%%%%%%%
{\bf NOTE:} In this paper $p(i,j)=1/deg(i)$ where $deg(i)$ is the degree of node $i$. However any  transition probabilities for which
the resulting Markov chain is ergodic can be used. \\

The matrix $\mathcal{P}=(p_{ij})_{i,j=1 \cdots N}$ is the transition matrix of a Markov chain which in our choice or any choice
of transition probabilities, is assumed to be irreducible and aperiodic (\cite{Kemeny}). Starting at any node outside of $A$, a random walker first reaches the set $A$ at a hitting time $T_{A}=\min\{n >0: X_{n} \in A \}$, where $X_n$ is the node occupied by the walker
at time $n$. The expected hitting time is $\mathbb{E}[T_{A}]$. If the walker starts at a fixed $i \notin A$, then the expected
hitting time is the conditional expectation $\mathbb{E}[T_{A}| X_{0}=i]=\mathbb{E}_{i}[T_{A}]$. Writing $h(i,A)=\mathbb{E}_{i}[T_{A}]$, the value of $F$ at $A$ is expressed as
%%%%%%%%%%%%%%%%%%
%
\begin{equation} \label{E:hitnumber}
F(A)={\sum}_{i \notin A}h(i,A).
\end{equation}
%%%%%%%%%%%%

Given $A$, $F(A)$ can be evaluated by solving a suitable linear equation. Indeed
a standard result in Markov chain theory \cite{Kemeny} tells us that $h(i,A)$ is the ith component of the vector $\mathsf{H}$,
which is the solution of the linear equation,
%%%%%%%%
%
\begin{equation}\label{E:lineq}
{\LARGE\mathsf{H}}=\bf{1}+\mathcal{P}_{A}{\LARGE\mathsf{H}},
\end{equation}
where $\mathbf{1}$ is a column vector of $N-|A|$ ones and  $\mathcal{P}_{A}$ is the matrix that results from
crossing out the rows and columns of $\mathcal{P}$ corresponding to the nodes of $A$. The value $F(A)$ is then the sum of the components
of $\LARGE\mathsf{H}$.
%------------------------------------------------------------------------------------------------------------------------------------

\subsection{Our Contribution and Organization of the Paper} \label{S:contribution}

We present an approach to the solution of the optimization problem (\ref{E:opt}) that generalizes  the classic greedy algorithm. In this paper we introduce a new
constraint set (optimal and near optimal sets) that fits within the framework of classes of sets that have the property of being closed under addition and judicious deletion of elements. Such classes e.g. matroids and greedoids play an important role in the optimization of modular and  sub(super) modular functions. Vertex covers are used to create optimal and near optimal sets and if high quality subsets exist, then the offered approximation is guaranteed to be better than the classic greedy algorithm. We present sufficient conditions for the performance ratio of our method to exceed the $(1-\frac{1}{e})$ ratio obtained by Borkar et al, and by Nemhauser et al for the greedy algorithm.
These results are to our knowledge new to the treatment of this problem and to the optimization of sub(super)modular functions generally.

The plan of the paper is as follows: section \ref{S:optnopt} contains a definition and discussion of optimal and near optimal sets ranked relative to a vertex cover of the graph $G$ with cardinality $C$. In sections \ref{S:maxmatch} and \ref{S:optnopt}, we demonstrate how the method is applied to a graph using a collection of sets $\mathbf{S}$ that are subsets of a vertex cover. If every vertex cover contained optimal sets as subsets, it would make sense to use this choice consistently. Unfortunately, optimality of a set is generally not preserved by the addition or deletion of nodes. 
We remedy this situation in part by introducing greedoids, a class of sets that are closed under the judicious deletion and addition of single elements. 

In section \ref{S:closure} we demonstrate the method on a second graph where $\mathbf{S}$  is chosen to be a group of feasible sets of a greedoid. 
In section \ref{S:quality}, the quality of the approximation obtained by our method is evaluated in terms of the ranking function $\bar{\rho}$ introduced in section \ref{S:optnopt}. A discussion of the computational complexity and tradeoff considerations can be found in section \ref{S:effort}. The statement and proof of the main result we described earlier is in section \ref{S:guarantee}. Our concluding discussion is found in section \ref{S:concl}.
%--------------------------------------------------------------------------------------------

\begin{figure}[tbp] % float placement: (h)ere, page (t)op, page (b)ottom, other (p)age
  \centering
  % file name: Z:/mydocs/BORKARpub/SEA-14/llncs2e/OptimalSeeds/pngs-for-OptRec/SVertexCover.png
  \includegraphics[width=5.36in,height=0.889in,keepaspectratio]{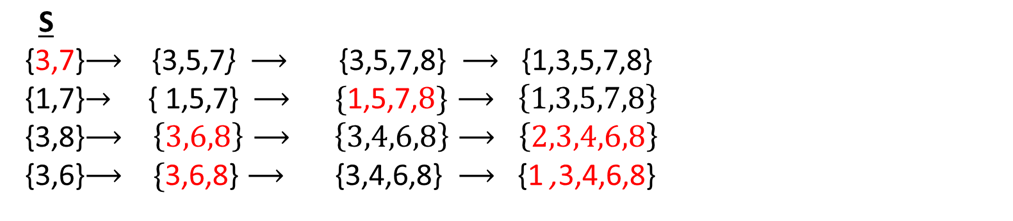}
  \caption{Optimal sets of the graph in Figure \ref{fig:JW8K1} for K=4 and 5 obtained by greedy extension of $\mathbf{S}$, subsets of a vertex cover}
  \label{fig:SVertexCover}
\end{figure}

%\begin{figure}[tbp] % float placement: (h)ere, page (t)op, page (b)ottom, other (p)age
%  \centering
%  % file name: Z:/mydocs/BORKARpub/IEEE_Trans_Network/IEEEtran/NetSci&Eng_figures/SVertexCover.eps
%  \includegraphics[bb=0 0 1013 221,width=4.5in,height=0.979in,keepaspectratio]{SVertexCover}
%  \caption{Optimal sets of the graph in Figure \ref{fig:JW8K1-jres} for K=4 and 5 obtained by greedy extension of $\mathbf{S}$, subsets of a vertex cover}
%  \label{fig:SVertexCover}
%\end{figure}
%
%----------------------------------------------------------------------------------------------

\begin{figure}[tbp] % float placement: (h)ere, page (t)op, page (b)ottom, other (p)age
  \centering
  % file name: Z:/mydocs/BORKARpub/SEA-14/llncs2e/OptimalSeeds/pngs-for-OptRec/JW8K1.PNG
  \includegraphics[width=3.41in,height=2.19in,keepaspectratio]{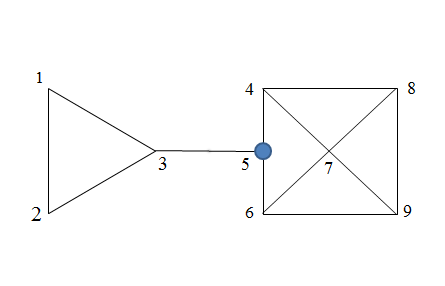}
  \caption{Graph with N=9 vertices, showing optimal set for K=1}
  \label{fig:JW8K1}
\end{figure}

%%%%%%%%%%%%%%%%%%%%%%%%%%%%%%%%%%%%%%%%%%%%%%%%%%%%%%%%%%%%%%%%%%%%%%%%%%%%%%%%%%%%%%%%%%%%%%%%%%%%%%%%%%%%%%%%%%%%%%%%%%%%%%%%%%%%%%
%%
\section{Finding and Approximating Optimal Sets} \label{S:newsolution}
\subsection{Maximal Matches}\label{S:maxmatch}

The optimization problem as posed in equation (\ref{E:opt}) assumes no advance knowledge about the optimal set or any
other possibly related sets. Thus let us first consider a process of obtaining optimal sets by using
subsets of existing ones.
Let $A$ be a vertex cover (not necessarily a minimum one). Since every edge is incident to a 
vertex of $A$, a random walker starting at a vertex $i$ outside of $A$ must hit $A$ at the first step. That is $h(i,A)=1$. Now equation (\ref{E:lineq}) implies that $h(i,A)\geq 1$ so it follows that $A$ must be an optimal set for its own cardinality. Thus a solution for $C=|A|$ is obtained by constructing a vertex cover. Fortunately a maximal match can be constructed by a simple greedy algorithm and its vertices are a vertex cover with cardinality $C \leq 2*(\mathcal{VC})$  where $\mathcal{VC}$ is the cardinality of a minimum vertex cover \cite{Cormen}. Therefore without loss of generality we turn our attention to the solution of problem (\ref{E:opt}) for $K \leq C$.
%------------------------------------------------------------------------------------------------------------------------------------
%
%
%
\subsection{Optimal and Near Optimal Sets}\label{S:optnopt}
We introduced a measure of the spread effectiveness of sets in equation (\ref{E:hitnumber}). It will be convenient to
convert this to a rank defined on subsets of $V$. In particular, suppose there exists a vertex cover with $C$ vertices.
We will order all non-empty subsets $A \subseteq V$ such that $|A|\leq C$ with a ranking function $\bar{\rho}(A)$ defined as,
\begin{equation}\label {E:rank}
\bar{\rho}(A)=\frac{F_{max}-F(A)}{F_{max}-F_{min}} 
\end{equation}
where $F_{max}=\max_{\emptyset \ne A\subseteq V, |A|\leq C}\large F(A)$, and $F_{min}$ is the corresponding minimum. $F_{min}$ can be calculated by computing $F$ for a vertex cover of cardinality $C$ whose elements might be e.g. the endpoints of a maximal match. We define $F_{max}$ to be the maximum value of $F$ among all one element subsets. We assume that $F_{max}\ne F_{min}$. If this were not the case, $F(A)$ would be have the same value for any non-empty subset $A$ with $|A| \leq C$. Thus any $A$ with $|A|\leq K$ would be a solution of the problem.

If $A$ is optimal and $|A|=C$ then $\bar{\rho}(A)=1$. Conversely, the worst performing set has value $0$.
For a constant $\nu \,( 0<\nu \leq 1)$ and $C$, the non-empty set
\begin{equation}\label{E:Lck}
\mathnormal{L}_{\nu,C}=\{A: A\subseteq V, |A|\leq C , \bar{\rho}(A)\geq \nu\}
\end{equation}
 defines a set of optimal and near optimal subsets, with the degree of near optimality depending on $\nu$. 
Let $m$ be the smallest cardinality of sets in $\mathnormal{L}_{\nu,C}$. 
Starting with a collection of sets $\mathbf{S} \subset \mathnormal{L}_{\nu,C}$ of size $m$, our method is to seek a solution 
to the optimization problem (\ref{E:opt}) by greedily augmenting each set until it reaches the desired size $K$. The offered approximation is the best (has the lowest $F$ value) of these extended sets. We can always find a $\nu$ and $C$ so that $\mathnormal{L}_{\nu,C}$ contains the optimal set of cardinality $K$ but we do not have a proof that the approximation generated by subsets of a vertex cover is optimal. However since our solution is a superset of sets in $\mathnormal{L}_{\nu,C}$, it is also in $\mathnormal{L}_{\nu,C}$ and therefore has minimum rank $\nu$. We illustrate the method with an example.

\begin{figure}[tbp] % float placement: (h)ere, page (t)op, page (b)ottom, other (p)age
  \centering
  % file name: Z:/mydocs/BORKARpub/SEA-14/llncs2e/OptimalSeeds/pngs-for-OptRec/JW8K2.PNG
  \includegraphics[width=3.41in,height=2.24in,keepaspectratio]{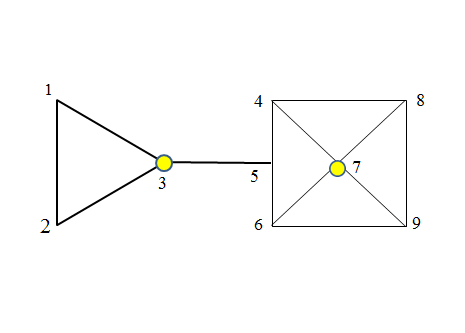}
  \caption{Optimal set K=2 for graph in Figure \ref{fig:JW8K1}}
  \label{fig:JW8K2}
\end{figure}

\begin{figure}[tbp] % float placement: (h)ere, page (t)op, page (b)ottom, other (p)age
  \centering
  % file name: Z:/mydocs/BORKARpub/SEA-14/llncs2e/OptimalSeeds/pngs-for-OptRec/JW8K3.PNG
  \includegraphics[width=3.41in,height=2.02in,keepaspectratio]{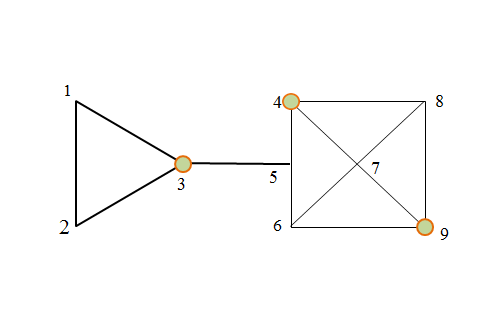}
  \caption{Optimal set K=3 for graph in Figure \ref{fig:JW8K1}. The set \{3,6,8\} is also optimal by symmetry }
  \label{fig:JW8K3}
\end{figure}

\begin{figure}[tbp] % float placement: (h)ere, page (t)op, page (b)ottom, other (p)age
  \centering
  % file name: Z:/mydocs/BORKARpub/SEA-14/llncs2e/OptimalSeeds/pngs-for-OptRec/JW8K4.PNG
  \includegraphics[width=3.41in,height=2.02in,keepaspectratio]{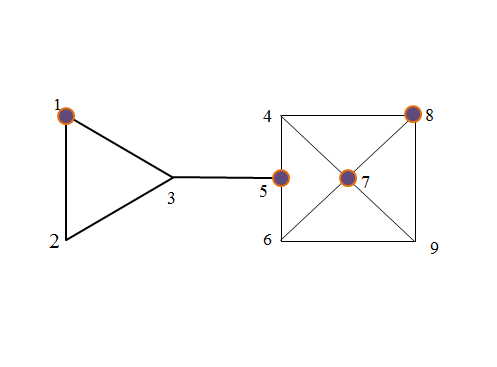}
  \caption{Optimal set K=4 for graph in Figure \ref{fig:JW8K1}}
  \label{fig:JW8K4}
\end{figure}

\begin{figure}[tbp] % float placement: (h)ere, page (t)op, page (b)ottom, other (p)age
  \centering
  % file name: Z:/mydocs/BORKARpub/SEA-14/llncs2e/OptimalSeeds/pngs-for-OptRec/JW8K5.PNG
  \includegraphics[width=3.41in,height=2.02in,keepaspectratio]{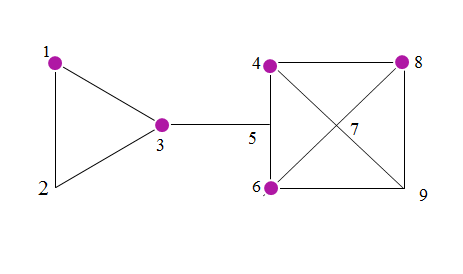}
  \caption{Optimal set K=5 for graph in Figure \ref{fig:JW8K1}}
  \label{fig:JW8K5}
\end{figure}

Figures \ref{fig:JW8K1} through \ref{fig:JW8K5} show a graph with $N=9$ 
vertices along with the vertices of optimal sets for $K=1$ through $5$.  To solve the problem for $K=4$, we note that the class of optimal and near optimal sets based on $C=8$ and $\nu=0.90$ has minimum set size $m=2$. The set $\mathcal{M}=\{1,3,5,6,7,8\}$  is a vertex cover (calculated from the maximal match algorithm). We define $\mathbf{S}$ to be the two element subsets of $\mathcal{M}$ that are 
in $\mathnormal{L}_{.90,8}$. The first column of Figure \ref{fig:SVertexCover} lists these sets and subsequent columns show the results of greedy one element extensions of $\mathbf{S}$ until $K=4$. Optimal sets are shown in red.  In this example the offered approximation is optimal. This is also the case for extensions up to $K=5$. In this case we see that the method identifies optimal sets that are subsets of $\mathcal{M}$ as well as others that are not, e.g., $\{2,3,4,6,8\}$, underlining the fact that the method finds sets that are reachable by greedy extension of subsets of $\mathcal{M}$. The offered approximation for this method is guaranteed to be in 
$\mathnormal{L}_{.90,8}$. This is a consequence of Proposition \ref{P:upextend} which is discussed and proved in Section \ref{S:closure}.\\

%--------------------------------------------------------------------------------------------------
%
\section{Closure Property of Optimal and Near Optimal Sets}\label{S:closure}
\begin{figure}[tbp] % float placement: (h)ere, page (t)op, page (b)ottom, other (p)age
  \centering
  % file name: Z:/mydocs/BORKARpub/SEA-14/llncs2e/OptimalSeeds/pngs-for-OptRec/BiggsK4B.PNG
  \includegraphics[width=3.49in,height=3.05in,keepaspectratio]{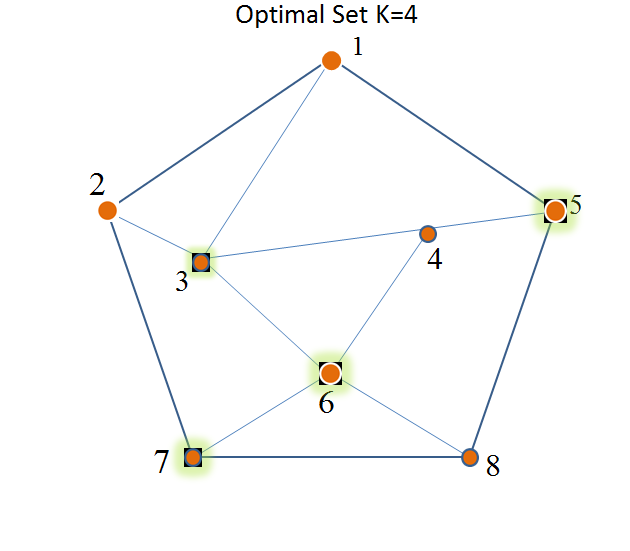}
  \caption{Graph with N=8, vertices. Vertices of optimal set K=4 shown as squares}
  \label{fig:BiggsK4B}
\end{figure}

\begin{figure}[tbp] % float placement: (h)ere, page (t)op, page (b)ottom, other (p)age
  \centering
  % file name: Z:/mydocs/BORKARpub/SEA-14/llncs2e/OptimalSeeds/pngs-for-OptRec/BiggsGREEDOID.PNG
  \includegraphics[width=3.92in,height=2.06in,keepaspectratio]{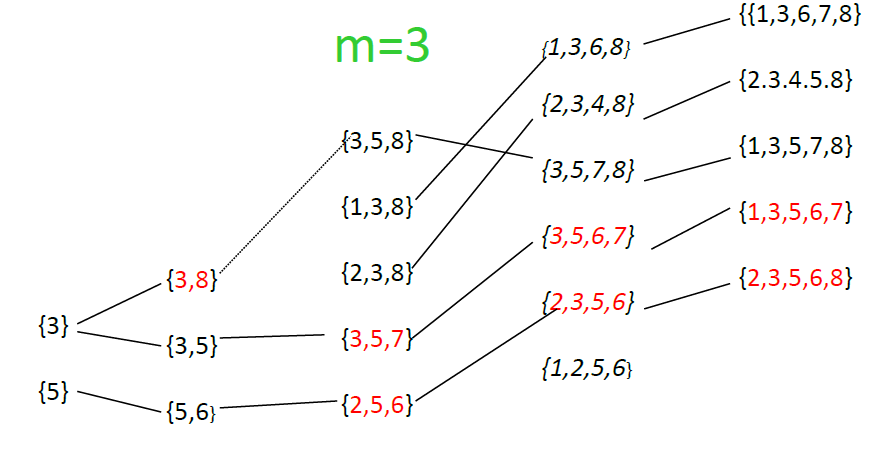}
  \caption{Greedoid constructed from optimal and near optimal sets $\mathnormal{L}_{.85,7}$ of graph in Fig \ref{fig:BiggsK4B}}
  \label{fig:BiggsGREEDOID}
\end{figure}

In section \ref{S:optnopt}, we demonstrated our method for approximating a solution of optimization problem  (\ref{E:opt}) based on greedy extensions of subsets of a vertex cover that are optimal or near optimal. Unfortunately a vertex cover can fail to have such subsets other than the vertex cover itself (see an example in \cite{Hunt1}). This is the motivation for finding other classes of optimal and near optimal sets that are closed under judicious addition and deletion of elements. We conjecture that greedy extension  of such sets will have the largest likelihood of success. The structure we seek is conveniently described in terms of a generalization of the matroid, known as a {\it greedoid}  \cite{Korte,Bjorner}.
\begin{Definition}\label{D:greedoid}
Let $\mathbf{E}$ be a set  and let $\mathcal{ F}$ be a collection of subsets of $\mathbf
{E}$. The pair $(\mathbf{E}, \mathcal{F})$  is called a \underline {greedoid} if $\mathcal{F}$ satisfies
\begin{itemize}
  \item $\mathbf{G1}:$ $\emptyset\in \mathcal{F}$ \label{null}
  \item $\mathbf{G2}:$ For $A \in \mathcal{F}$ non-empty, there exists an $a \in A$ such that $A \setminus \{a\} \in \mathcal{F}$ \label{access}
  \item $\mathbf{G3}:$ Given $X$, $Y$ $\in \mathcal{F}$ with $|X| > |Y|$, there exists an $x \in X\setminus Y$, such that $Y \cup \{x\} \in \mathcal{F}$ \label{augment}
\end{itemize}
\end{Definition}
A set in $\mathcal{F}$ is called \underline {feasible.} Note that $\mathbf{G2}$ implies that a  single element can be removed from a feasible set $X$ so that the reduced set is still feasible. By repeating this process the empty set eventually is reached. Conversely
starting from the empty set, $X$ can be built up in steps through repeated use of $\mathbf{G3}$.\\
We now show that $\mathnormal{L_{\nu,C}}$ satisfies condition $\mathbf{G3}$ of the definition
for any  $0 < \nu \leq 1$,\, $0\leq C \leq N$ (Proposition \ref{P:upextend}). The proof depends on the following lemma and uses
an adaptation of an argument in Clark et al \cite{Clark}
\begin{lemma}\label{L:mono}
Let $S \subseteq V$, $u \in V\setminus S$. Then $F(S) \geq F(S\cup \{u\})$. 
\end{lemma}
{\bf Proof:} Suppose $S$-a set of nodes, is a target set for a random walk. Let $E_{ij}^{l}(S)$ be the event,
$E_{ij}^{l}(S)=\{ X_{0}=i \in V,\   X_{l}=j \in V\setminus S,\ X_{r} \notin S,\ 0 \leq r \leq l \}$.
Thus paths of the random walk in this event start at $i$ and arrive at $j$ without visiting $S$ during the interval $[ 0 , l ]$. 
Also define the event $F_{ij}^{l}(S,u)=E_{ij}^{l}(S) \cap \bigcup_{m=0}^{l} \{ X(m)=u\}$ where $u \notin S$. 
Paths in this event also start at $i$ and arrive at $j$ without visiting $S$, but  must visit the element $u$ at some time during the interval $[ 0 , l ]$. Since a path either visits $u$ in the time interval $[ 0 , l ]$ or it does not, it follows that:
\begin{equation}\label{E:seteq}
E_{ij}^{l}(S)=E_{ij}^{l}(S \cup \{u \})\cup F_{ij}^{l}(S,u)
\end{equation}
We have  $E_{ij}^{l}(S \cup \{u\}) \bigcap F_{ij}^{l}(S,u)=\emptyset$. This implies that,
\begin{equation}\label{E:indeq}
\mathbf{1}_{E_{ij}^{l}(S)}=\mathbf{1}_{E_{ij}^{l}(S \cup \{u\})}+\mathbf{1}_{F_{ij}^{l}(S,u)}
\end{equation}
and therefore:
\begin{equation}\label{E:notindeq}
\mathbf{1}_{E_{ij}^{l}(S)} \geq \mathbf{1}_{E_{ij}^{l}(S \cup \{u\}}
\end{equation}
Here $\mathbf{1}_{A}$ is the usual indicator function of the set $A$, i.e. the function $\mathbf{1}_{A}: \Omega \rightarrow \{ 0, 1\}$ .
Recalling that $T_{S}$ is the hitting time for
set $S$, the following relation comes from taking the expection of $\mathbf{1}_{E_{ij}^{l}(S)}$ on the left hand side of (\ref{E:notindeq}) summing over all $j \in V\setminus S$. Here $\mathbb{E}$ denotes expectation.
\begin{equation}\label{E:hitS}
\mathbf{Prob}\{ T_{S} > l | X_{0}=i\}=\mathbb{E}\left(\sum_{j \in V \setminus S}\mathbf{1}_{E_{ij}^{l}(S)}\right)
\end{equation}
A similar result is obtained for $T_{S\cup \{u\}}$ from taking the expectation of $\mathbf{1}_{E_{ij}^{l}(S\cup\{u\})}$ on the right
hand side of (\ref{E:notindeq}) and summing over $j \in V\setminus S$.
Summing once again over all $l \geq 1$ results in the inequality,
\begin{equation}\label{E:hitsumineq}
h(i,S) \geq h(i,S\cup \{u\})
\end{equation}\\
$\Box$ \\
\textbf{REMARK:} See (section 5 in \cite{Hunt2}), for an explicit formula for the increment $F(S)-F(S\cup\{u\})$.
\begin{proposition}\label{P:upextend}
For $0 < \nu \leq 1$ and $0 < C \leq N$, let $\mathnormal{L}_{\nu,C}$ be the class of sets defined in equation (\ref{E:Lck}).
Then $\mathnormal{L}_{\nu,C}$ satisfies condition  $\mathbf{G3}$.
\end{proposition}
{\bf Proof:} The conclusion follows from the definition of $\mathnormal{L}_{\nu,C}$ and Lemma \ref{L:mono}.
$\Box$  \\ \\
The proposition establishes that $L_{\nu,C}$ satisfies the $\mathbf{G3}$ property for greedoids. However, $\mathbf{G2}$ does not hold. For example
if the set $A$ has cardinality $m$ where $m$ is the size of the smallest set in $L_{\nu,C}$ then $A\setminus \{a\}$ cannot be in 
$\mathnormal{L_{\nu,C}}$  for any element $a \in A$. Conversely, let $c_{n}=\max_{|X| \leq n}\rho(X)$. If $c_{m}\geq c>c_{m-1}$ then $m$ is the size of the smallest set in $\mathnormal{L_{\nu,C}}$. Define $G_n$ to be all sets in $\mathnormal{L}_{\nu,C}$ of cardinality $n$. To create a class of sets with the $\mathbf{G2}$ property, one constructs subsets of $G_m$ of size $n \leq m$ that  are "augmentable", i.e., that satisfy $\mathbf{G3}$. Sets $G_n$ for $n > m$ are culled so the remaining sets are supersets of the "augmentable" sets and therefore satisfy $\mathbf{G2}$. The greedoid will then consist of selected subsets and supersets of $G_m$. Conditions for the existence of "augmentable" subsets of $G_m$ and proof of the validity of the resulting greedoid construction can be found in \cite{Hunt1}. Rather than repeat the details of these arguments here, we close this section with an example showing the greedoid of a graph (Figure \ref{fig:BiggsK4B}) and its use in the solution of (\ref{E:opt}).The minimum cardinality of a set in the class of optimal and near optimal sets $\mathnormal{L}_{.85,7}$ is $m=3$. These sets are used to create the greedoid depicted in Figure \ref{fig:BiggsGREEDOID}. Note that $\mathbf{G1}$-$\mathbf{G3}$ are satisfied.
Assume the optimal set for $K=4$ is unknown. Then our method in this case is to take $\mathbf{S}$ to be the three element sets in 
$\mathnormal{L}_{.85,7}$  that are feasible sets of the greedoid and perform a greedy extension of each set. In Figure
 \ref{fig:BiggsGREEDOID} a line is drawn between a set and its greedy extension. We have also drawn greedy extensions of sets of cardinality $n < m$ as well.
The optimal sets are shown in red and so they are in the greedoid. The offered approximations are in fact exact. This second general approach of using the starter set $\mathbf{S}$ to be a collection of feasible sets of a greedoid contains the vector cover based approach as a special case. Note that the the vector cover subsets in the starter set are restricted to be optimal and near optimal sets in $L_{\nu, C}$ rather than arbitrary subsets. Thus they must have
a minimum quality $\nu$ and have cardinality less than $C$.

The methods we present here are similar to some earlier approaches to the problem. For example the enumeration technique for maximizing a submodular function subject to a modular function constraint  (\cite{Sahni},\cite{Khuller}, \cite{Sviridenko}). Rather than begin the greedy algorithm with an optimal singleton, the enumeration technique performs a greedy extension of all k element subsets (where $k \leq m=3$). Khuller et al (\cite{Khuller}) and then Sviridenko (\cite{Sviridenko}) proved these methods have a $(1-\frac{1}{e}])$ performance guarantee and Feige in (\cite{Feige}) proved this guarantee was the best possible. In our setting, $m$ is associated with a lower bound of the rank of near optimal sets $\nu$. In section \ref {S:guarantee} we show how an improvement in the performance guarantee is possible if the rank of the greedy solution is less than some $\eta$ where $\eta \geq \nu$. 
Our method also has connections to the MMax the ascent algorithm for constrained maximization of a monotone submodular function, applied to $-F$  ( see section 6.2 in \cite{Iyer1}).  We iterate a greedy subgradient equation for  each member of the starting set $\mathbf{S}$. Successive iterates are non-decreasing and converge to an extreme point of the subdifferential of $-F$ at the $K$ element terminal set containing the initial set(\cite{Iyer2}). 
%%
%%%%%%%%%%%%%%%%%%%%%%%%%%%%%%%%%%%%%%%%%%%%%%%%%%%%%%%%%%%%%%%%%%%%%%%%%%%%%%%%%%%%%%%%%%%%%%%%%%%%%%%%%%%%%%%%%%%%%%%%%%%%%%%%%%%%%%

%-------------------------------------------------------------------------------------------

%---------------------------------------------------------------------------------------------

%--------------------------------------------------------------------------------------------------

%----------------------------------------------------------------------------------------------

%----------------------------------------------------------------------------

%
%
%%%%%%%%%%%%%%%%%%%%%%%%%%%%%%%%%%%%%%%%%%%%%%%%%%%%%%%%%%%%%%%%%%%%%%%%%%%%%%%%%%%%%%%%%%%%%%%%%%%%%%%%%%%%%%%%%%%%%%%%%%%%%%%%%%%%%%
%%
%%
%%
%%
\section{Quality of the approximation}\label{S:quality}
%---------------------------------------------------------------------------------------------------------------------------------
\subsection{Comparison with the optimal solution and greedy solution} \label{S:submodcompare}

Following Ilev (\cite{Ilev}), $F$ can be defined for the empty set as 
\begin{equation}
0\leq F(\emptyset)=\max_{X \cap Y=\emptyset , X, Y \subseteq V} F(X) +F(Y)-F(X \cup Y) < \infty
\end{equation}
Thus by the definition of $\bar{\rho} \, \,$,  $\bar{\rho}(\emptyset)=\frac{F_{max}-F(\emptyset)}{F_{max}-F_{min}}$.
This means the normalized function defined on sets $A , \,$ $\rho(A)=\bar{\rho}(A)-\bar{\rho}(\emptyset)$ is bounded, submodular and non-decreasing. Also note that since $F(\emptyset) \geq F_{max}$,  $\rho(A) \geq 0$ for all $A \subseteq V$. For the empty set we have $\rho(\emptyset)=0$. For the remainder of this paper, we will refer to $\rho$ as the normalized rank function and to $\bar{\rho}$ as the un-normalized rank function. Since the rank is an affine function of $F$, the optimization problem (\ref{E:opt}), is equivalent to the problem of finding the set that  maximizes the normalized rank subject to the same constraints. This is a special case of the general problem first considered by Nemhauser, Wolsey and Fisher
\cite{NemWolFish}.

Our offered solution is the result of the greedy extension of a group of optimal and near optimal sets of minimal cardinality $m$. In this section we will compare the solution to the optimal solution using the normalized rank function $\rho$. Specifically, we show that the inequality of Nemhauser, Wolsey and Fisher, (\cite{NemWolFish}, Section 4) holds. Recall that they proved that the ratio of the normalized rank of the classic greedy solution to the optimal solution has
a lower bound of $(1-\frac{1}{e})$.

Let $S_{g}^{(m)}$ be the $m$ element set that is the result of greedily adding single elements $m$ times. We first suppose that $S_{g}^{(m)} \in \mathbf{S}$. 
\begin{lemma} \label{L:NWF} Suppose $S_{g}^{(m)} \in \mathbf{S} \subseteq \mathnormal{L}_{\nu,C}$. Let $S_{g}$ be the K element set obtained from the greedy extension of $S_{g}$.  If $S^{*}$ is the offered solution, then
\begin{equation}
F(S^{*}) \leq F(S_{g})
\end{equation}
\end{lemma}
{\bf Proof:} $F(S^{*})$ is the minimum value of all the values obtained by the greedy $K-m$ extension of elements in $\mathbf{S}$.
$\Box$ \\
The set $S_{g}$ is also the result of greedily adding single elements $K$ times. Thus we may use section 4 in \cite{NemWolFish} and the definition of $\rho$ to conclude that
\begin{corollary}\label{C:bound}
If $S^{*}$ is the solution constructed by the method described in sections \ref{S:optnopt} and \ref{S:closure} above, then 
\begin{equation}
\rho(S^{*})\geq (1-\frac{1}{e})\rho(\mathcal{O}_{K}^{*})
\end{equation}
where $\mathcal{O}^{*}_{K}$ is the solution of the optimization problem.\\
\end{corollary}

Once $F(S^{*})$ and $F(S_{g})$ have been computed we can determine $\chi$ such that 
$\rho(S^{*})=(1+\chi)\rho(S_{g})$. The quantity $\chi$ measures the degree of improvement of
$F(S^{*})$ over $F(S_{g})$. In particular we have, \\ 
\begin{proposition}  \label{P:betterbound} When $F(S^{*})<F(S_{g})$, so $\chi >0$, then
\begin{equation}
\rho(S^{*})\geq  (1+\chi)(1-\frac{1}{e})\rho(\mathcal{O}^{*}_{K}) \label{E:optimalcompare} \\
\end{equation}
\end{proposition}
\textbf{NOTE:} If $m=1$, then optimal and near optimal singletons of quality $\nu$
would be in $L_{\nu,C}$. Thus  necessarily $S^{(m)}_{g}$ must be in $\mathbf{S}$. In other
words for $m=1$, our method is a direct generalization of the classic greedy method.\\
 
If $S_{g}^{(m)} \notin \mathbf{S}$, then it may be possible to identify  a set $S$
 with $\rho(S) > \rho(S_{g})$ and $|S| \leq K$. In section \ref{S:guarantee} we discuss how this can be done when bounds on the curvature of $\rho$ and increments of $F$ are available. The existence of such an $S$ allows us to formulate sufficient conditions for $S^{*}$ to satisfy the hypothesis of Proposition \ref{P:betterbound}.

A lower bound of $(1-\frac{1}{e})$ was established by Borkar et al in \cite{Borkar}. 
Specifically it is a lower  bound on the ratio of $F(S_g)-F(\{a\})$ 
to $F(\mathcal{O}^{*}_{k})-F(\{a\})$, where $S_g$ is the result of the greedy algorithm starting with singleton $a$. 
%------------------------------------------------------------------------------------------------------------
%
\subsection{Computational effort and tradeoff with quality} \label{S:effort}
A rough estimate of the complexity of the method follows from realizing that the collection $\mathbf{S} \in$ 
$\mathnormal{L}_{\nu,C}$, has at most $\binom{N}{m},\,$  $m$ element sets. To determine whether or not a particular set is near optimal,
equation(\ref{E:lineq}) must be solved and this involves $O(N^{3})$ operations. Thus the objective function values of  $\mathbf{S}$ are determined in 
$O(N^{m+3})$ operations. The greedy extension of an $m$ element to a $K$ element set involves $O((K-m)(N-m))=O(N^{2})$ so that the extension of every set in $\mathbf{S}$ involves $O(N^{m+2})$ operations. Overall then, 
the method requires $O(N^{m+3})+O(N^{m+2})=O(N^{m+3})$ operations.
Thus it is desirable to make $m$ as small as possible. In fact we assume $m \ll K$. However the size of $m$ affects the accuracy.
%--
 Taking $\nu$ to be a measure of the quality of the approximation, we want to know given $m$,what $\nu$ can be expected?
Conversely given a desired quality $\nu$, what $m$ is required?
 We will employ the forward elemental curvature of the normalized rank function. Elemental curvature was used by Wang, Moran, Wang, and Pan \cite{Wang} in their treatment of the problem of maximizing a monotone non-decreasing submodular function subject to a matroid constraint. 

The elemental curvature of $\rho$ is defined 
over $\mathnormal{L}_{\nu,C}$ in terms of the marginal increase in the rank of a set when a single element is added to it.
First let $A$ be a set and $i \notin A$,
\begin{equation} \label{E:rankincr} 
\rho_i(A)=\rho(A \cup i)-\rho(A).
\end{equation}
and then for a fixed $A \in \mathnormal{L}_{\nu,C}$ set,
\begin{equation} \label{E:curveS}
k_{ij}(A)=\frac{\rho_{i}(A \cup j)}{\rho_{i}(A)}. 
\end{equation}
The curvature is defined then as,
\begin{equation} \label{E:curvature}
\kappa=\max\{ k_{ij}(A) : A \subset \mathnormal{L}_{\nu,C}\,\, , i\ne j,\, i, j \notin A\} .
\end{equation}

Since $\rho$ is submodular  $\kappa \leq 1$. 
Now suppose $S \subset T \subset \mathnormal{L}_{\nu,C}$. Given $\nu$, we want to determine the minimum size
of $S$ for which $\rho(S) \geq \nu$. 
If $T \setminus S=\{j_1, \cdots j_{r} \}$ , we have (see equation (2) in  \cite{Wang}) ,
\begin{equation} \label{E:setdiff}
\bar{\rho}(T)-\bar{\rho}(S)=\rho(T)-\rho(S)=\sum_{t=1}^{r}\rho_{j_{t}}(S \cup \{j_1 , \cdots j_{t-1}\}).
\end{equation}
Therefore ,
\begin{equation} \label{E:rankineq}
\bar{\rho}(T)-\bar{\rho}(S) \leq \rho_{j_1}(S)+\kappa\rho_{j_2}(S)+ \cdots \kappa^{t-1}\rho_{j_{r}}(S)
\end{equation}
Suppose $\bar{\rho}(T)=1$, for example if $T$ is a vertex cover. Define $\gamma$ to be $\gamma=\max\{\rho_{j_{t}}(S): S \subset T,\, t=1 \cdots r\}$.
 We can get a lower bound on the rank of $S$ using equation (\ref{E:rankineq}) and the inequality $0 \leq \rho_{j}(S)\leq \gamma$. First assume $\gamma$ is known. We know that if $S \ne \emptyset$, then $\gamma < 1$. Then,
\begin{equation} \label{E:ranklower}
\bar{\rho}(S) \geq 1-\gamma\sum_{t=1}^{r}\kappa^{t-1}
\end{equation}
Let us now suppose that :
\begin{equation}\label{E:rank&nu}
 \bar{\rho}(S) \geq \bar{\eta}=(1- \gamma\sum_{t=1}^{r}\kappa^{t-1})\geq \nu ,
\end{equation}
and $|S|\geq m$. If an approximation with quality $\nu$ is required, and
$r(\nu)$ is the largest value of $r$ such that inequality (\ref{E:rank&nu}) holds, then $r  \leq r(\nu)$. Now $K=C-r$ is the
cardinality of $S$ so that $C-r(\nu) \leq C-r$. Thus the smallest possible value of $|S|$ is
\begin{equation}
m(\nu)=C-r(\nu)
\end{equation}
In particular any $m$ must satisfy $m \geq m(\nu)$.
Conversely, given $m$, the quality of the approximation depends on $\gamma$ and $r=C-m$.  
More precisely, the largest value of $\nu$ and thus the largest guaranteed quality of an approximation obtained by our method, has an upper bound given by the right hand side of inequality (\ref{E:ranklower}). 

%%%%%%%%%%%%%%%%%%%%%%%%%%%%%%%%%%%%%%%%%%%%%%%%%%%%%%%%%%%%%%%%%%%%%%%%%%%%%%%%%%%%%%%%%%%%%%%%%%%%%
%
\subsection{When is our approximation is better than the greedy solution?}\label{S:guarantee}

Clearly if $F(S^{*})$ is available then we can check that $F(S^{*}) < F(S_{g})$.
However the value of $F(S^{*})$ is needed to calculate the degree of improvement. If additional knowledge of the graph, expressed as information  about $F$ is available an a priori estimate of the degree of improvement is possible without calculating $F(S^{*})$.
When $T$ is a vertex cover we obtained a lower bound on the rank of $S \subseteq T$  in terms of upper bounds on the increments and curvature of $\rho$ and the rank of $T$. With this bound in hand we would like to sharpen the comparison between the rank of the set approximation $S^{*}$ obtained by our method and the rank of the optimal set. We will derive a lower bound on the constant $\chi$. To do this, notice that if the un-normalized rank of $S^{*}$ satisfies $\bar{\rho}(S^{*})>\bar{\eta}$ and the greedy solution for the same cardinality satisfies
 $\bar{\rho}(S_{g}) \leq \bar{\eta}$, the normalized ranks of the respective sets satisfy 
$\rho(S^{*})>\eta$, and $\rho(S_{g})\leq \eta$ , where $\eta=\bar{\eta}-\bar{\rho}(\emptyset)$. Now as in section \ref{S:submodcompare}, $\rho(S^{*})=(1+\chi)\rho(S_{g})$ . So if we write $\rho(S_{g})=(1-\delta)\eta$ for some 
$0 <\delta<1$, then $\rho(S^{*})>\eta$ implies that $1+\chi>\frac{1}{1-\delta}$.
Thus $\chi > \frac{\delta}{1-\delta}$.

Rather than $S^{*}$ we could easily suppose there exists some $S$ that is a $K$ element greedy extension of a set in $\mathbf{S}$ such that $\rho(S) > \eta \geq \nu$. Arguing as before can conclude that 
$\rho(S)=(1+\chi)\rho(S_{g})$ for some $\chi > \frac{\delta}{1-\delta}$. Since $\rho(S^{*}) \geq
\rho(S)$, then we can still conclude that $\chi$ in inequality(\ref{E:optimalcompare})  has the claimed lower bound. Thus in summary, information about the the curvature and increments of $\bar{\rho}$ (equivalently $\rho$), can be used to obtain a lower bound $\eta>0$ on the rank of a set $S$ of cardinality $K$ when $S \subseteq T$, a vertex cover. Let $S$ also be the greedy extension of an element in $\mathbf{S}$, in other words, it is a set that arises from the implementation of our method. For example $\mathbf{S}$ could be collection of subsets of a vertex cover $T$. Alternatively, $\mathbf{S}$ could be a more general class of $m$ element feasible sets of a greedoid. Greedy extensions of such sets can be subsets of a vertex cover (see the implementation for the graphs in this paper for example). If $\rho(S_{g}) < \eta$ then inequality (\ref{E:optimalcompare}) holds with $\chi > \frac{\delta}{1-\delta}$. Thus we have a direct comparison of our approximation $S^{*}$ with the optimal solution. The degree of improvement over the classic greedy method, i.e. $\chi$ depends on the greedy solution
$S_{g}$, $\eta$, $\nu$ and $T$. This is the content of the following proposition.
%------------------------------------------------------------------------------------------------------------

\begin{proposition}\label{P:main}  Assume the following conditions:
\begin{enumerate}
  \item Let $T$ be a vertex cover and for $0 < \nu <1$, let $\mathbf{L}_{\nu,C}$ be a corresponding class of optimal and near optimal sets. Let $S \subseteq T$ be a $K$ element set that is the greedy extension of a set in $\mathbf{S}\subset \mathbf{L}_{\nu,C}$.
  \item $\rho(S)>\eta > 0$ where $\eta=\bar{\eta}-\bar{\rho}(\emptyset)$ and $\bar{\eta}$ is defined in inequality(\ref{E:rank&nu}).
  \item $S_{g}$ is a $K$ element set resulting from the greedy method. The rank of $S_{g}$ satisfies, $\rho(S_{g})<\min(\nu,\eta)$
\end{enumerate}
Then the offered approximation $S^{*}$ satisfies 
\begin{equation} \label{E:improved-opt}
\rho(S^{*})\geq (1+\chi)(1-1/e)\rho \left(\mathcal{O}_{K}^{*}\right) 
\end{equation}
where $\mathcal{O}_{K}^{*}$ is the optimal solution of problem (\ref{E:opt}) and for some $\delta$, $0 < \delta <1$,
 $\chi$ satisfies, 
\begin{eqnarray} \label{E:chi-bound}
\chi > \frac{\delta}{(1-\delta}.
\end{eqnarray}
We may set  $1-\delta=\frac{\rho(S_{g})}{\eta}$ so that $\delta$ depends on $S_{g}$, $\eta$,  $\nu$ and $T$.
\end{proposition}
%
%
%%%%%%%%%%%%%%%%%%%%%%%%%%%%%%%%%%%%%%%%%%%%%%%%%%%%%%%%%%%%%%%%%%%%%%%%%%%%%%%%%%%%%%%%%%%%%%%%%%%%%%%%%%%%%%%%%%%%%%%%%%%%%%%%%%%%%%%

\textbf{REMARK:} Our discussion suggests $S$ might be found by applying the backward greedy (worst-out) algorithm to a vertex cover with $T :\, |T|>K $ until a set of cardinality $K$ is reached. The conditions of the proposition can then be checked on the resulting $S$. If $S$ has a higher rank than the greedy solution but is not the greedy extension of a set in $\mathbf{S}$, then a better approximation or solution of the problem may result from  a series of comparisons  or single element exchanges between $S^{*}$ and $S$. \\

In this section we stated conditions that imply the existence of a sets $S$ whose rank (as expressed in terms of the elemental forward curvature of $\rho$), exceeds the rank of the greedy solution. If $S$ is also a greedy extension of a set in $\mathbf{S}$ then $S^{*}$ our offered solution has a higher degree of optimality as expressed in inequalities (\ref{E:improved-opt}) and (\ref{E:chi-bound})

%%%%%%%%%%%%%%%%%%%%%%%%%%%%%%%%%%%%%%%%%%%%%%%%%%%%%%%%%%%%%%%%%%%%%%%%%%%%%%%%%%%%%%%%%%%%%%%%%%%

\section{Conclusion}\label{S:concl} 
We posed the problem of identifying a subset of nodes in a network that will enable the fastest spread of consensus in a  decentralized communication environment. In a model of communication based on a random walk on an undirected graph $G=(V,E)$, the optimal set of nodes is found by minimizing the sum of the mean times of first arrival to the set by walkers who start at nodes outside the set.
Since the objective function for this problem is supermodular, the greedy algorithm has been a principal method for constructing approximations to optimal sets. Previous results guarantee that these sets are in some sense within $(1-1/e)$ of optimality.

   In this work we took a different approach. Rather than consider the problem (\ref{E:opt}) over all subsets of cardinality up to $K$, we restricted the search for an optimizing set to classes of optimal and near optimal sets that are closed under the addition and judicious deletion of elements. These sets have a predefined degree of near optimality (see section \ref{S:optnopt}). Let $m$ be the minimum cardinality of sets in this class.
We offered an approximation of the solution of optimization problem based on the greedy extension of a starter set $\mathbf{S}$ of sets of size $m$. In actual implementation we take $m \leq 3$ and theoretical arguments suggest this is a good choice. We demonstrated the method for two 
choices of $\mathbf{S}$, first a class subsets of a vertex cover and then for the feasible sets of a greedoid constructed from $\mathnormal{L}_{\nu,C}$. Here $\nu$ measures the degree of near optimality relative to a vertex cover of cardinality $C$. As shown in section 
\ref{S:submodcompare}, when the greedy solution at stage $m$ is in the intial set $\mathbf{S}$ then the results of our method are (unsurprisingly) at least as good.
It is also clear that if $\mathbf{S}$ has a greedy extension that contains a set $S, |S| \leq K$, that is better than the greedy solution of problem (\ref{E:opt}), our offered approximation $S^{*}$must also be better. Our principal result is a set of sufficient conditions and a proof that they imply the existence of such a set 
(see section \ref{S:guarantee}). It is a high ranking subset of a vertex cover and its rank can be estimated when information about the elemental forward curvature and increments of $F$ are available. Moreover $S^{*}$ can be directly compared with optimal solution and a lower bound on the improvement in optimality can be obtained without explicit knowledge of $S^{*}$.

In closing, we propose that the method we describe here is an instance of a more general approach based on the navigation in a graph whose nodes are optimal and near optimal sets. In this paper, the search for an optimal solution is based on greedily moving forward. However the most effective approach may be a more general class of steps. Movement between adjacent nodes would correspond to the addition (forward) and deletion (backward) or exchange of single elements.
%%%%%%%%%%%%%%%%%%%%%%%%%%%%%%%%%%%%%%%%%%%%%%%%%%%%%%%%%%%%%%%%%%%%%%%%%%%%%%%%%%%%%%%%%%%%%%%%%%%
%
%
%
%
%\textbf{Acknowledgement:} The author thanks Jason Wu of the University of California,
%Berkeley for his  assistance in the initial stages of this project.

%
%

%
%
% Set the ending of a LaTeX document
\end{document}